\documentclass[smallabstract,smallcaptions]{dccpaper}
\usepackage{epsfig}
\usepackage{citesort}
\usepackage{amsmath}
\usepackage{amssymb}
\usepackage{color}
\usepackage{url}
\usepackage{amsthm}
\usepackage{xspace}

\usepackage{graphicx} % omit "demo" for your real document
\usepackage{subcaption}

\usepackage{array}

\usepackage{siunitx}

\usepackage{tabularx}

\def\ie{{\em i.e.}\xspace}

\usepackage[outdir=./]{epstopdf}

\allowdisplaybreaks

\begin{document}
\title{\large \textbf{On dynamic succinct graph representations}}
\author{%
Miguel E. Coimbra$^{\ast}$,
Alexandre P. Francisco$^{\ast}$,
Lu\'{i}s M. S. Russo$^{\ast}$,\\
Guillermo de Bernardo$^{\dag,\S}$,
Susana Ladra$^{\dag}$,
Gonzalo Navarro$^{\ddag}$\\[0.5em]
{\small\begin{minipage}{\linewidth}\begin{center}
\begin{tabular}{cc}
$^{\ast}$INESC-ID / IST           & $^{\dag}$Universidade da Coru\~{n}a  \\
Universidade de Lisboa            & Centro de investigación CITIC                                \\
Portugal                          & Facultad de Informática                               \\
\url{miguel.e.coimbra@tecnico.ulisboa.pt} & Spain             \\
\url{aplf@tecnico.ulisboa.pt}             & \url{gdebernardo@udc.es}                 \\
\url{luis.russo@tecnico.ulisboa.pt}       & \url{sladra@udc.es} \\[0.5em]
$^{\S}$Enxenio SL & $^{\ddag}$IMFD, Dept. of Computer Science \\
Spain & University of Chile \\
\url{gdebernardo@enxenio.es} & Chile \\
& \url{gnavarro@dcc.uchile.cl}
\end{tabular}
\end{center}\end{minipage}}
}
\maketitle              % typeset the title of the contribution
\thispagestyle{empty}
\begin{abstract}
We address the problem of representing dynamic graphs using $k^2$-trees.
The $k^2$-tree data structure is one of the succinct data structures proposed for representing static graphs, and binary relations in general.
It relies on compact representations of bit vectors.
Hence, by relying on compact representations of dynamic bit vectors, we can also represent dynamic graphs.
%This approach suffers however of a well known bottleneck in compressed dynamic indexing.
In this paper we follow instead the ideas by Munro {\em et al.}, %~\cite{DBLP:conf/pods/MunroNV15}, %to circumvent this bottleneck and obtain a practical implementation.
and we present an alternative implementation for representing dynamic graphs using $k^2$-trees.
Our experimental results show that this new implementation is competitive in practice.
\end{abstract}
\Section{Introduction}\label{sec:introduction}
%
%Motivation
%
Graphs are ubiquitous among many complex systems, where we find large and dynamic complex networks.
It is therefore important to be able to not only store such graphs in compressed form, but also to update and query them efficiently while compressed.
Most succinct data structures for representing graphs are however static~\cite{DBLP:conf/www/BoldiV04,Brisaboa2014}.
And only recently, by relying on compact representations of dynamic bit vectors, succinct representations for dynamic graphs were presented~\cite{DBLP:journals/is/BrisaboaCBN17}.
These representations suffer however from a well known bottleneck in compressed dynamic indexing~\cite{DBLP:conf/pods/MunroNV15,navarro2016compact}.
In this paper we adopt the ideas proposed by Munro {\em et al.}~\cite{DBLP:conf/pods/MunroNV15} to represent dynamic graphs through collections of static and compact graph representations.

%Contribution
%
Our approach relies on $k^2$-trees to represent static graphs and our implementation supports both edge insertions and deletions with almost the same (but amortized) cost as static $k^2$-trees.
We provide an implementation and an extensive experimental evaluation.

% NOTE: this \section command was commented out because of the following LaTeX warning in miktex:
% Package hyperref Warning: Token not allowed in a PDF string (PDFDocEncoding): removing `math shift' on input line 69.
% Package hyperref Warning: Token not allowed in a PDF string (PDFDocEncoding): removing `superscript' on input line 69.
% Package hyperref Warning: Token not allowed in a PDF string (PDFDocEncoding): removing `math shift' on input line 69.
%\section{From static $k^2$-trees to dynamic graphs}~\label{sec:static-to-dynamic}
\Section{From static $k^2$-trees to dynamic graphs}\label{sec:static-to-dynamic}
%
%Main idea
%
Let $G=(V,E)$ be a graph where $V$ is the set of vertices, with size $n$, and $E\subseteq V\times V$ is the set of edges, with size $m$.
The main idea is to represent $G$ dynamically, supporting edge insertions and deletions, as well as common operations over graphs, through a collection of static edge sets $\mathcal{C}=\{E_0,\ldots,E_r\}$.
Each static edge set $E_i$ is then represented using a static $k^2$-tree, except $E_0$ which is represented through a dynamic and uncompressed adjacency list.

As discussed by Munro {\em et al.}~\cite{DBLP:conf/pods/MunroNV15}, we must control both the number of edges $m_i$ in each set $E_i$ and the number $r$ of such sets to achieve the optimal amortized cost for each operation.
The first set ($E_0$) contains at most $m/\log^2 m$ elements.
In general we require that $m_i$ is at most $m/\log^{2 - i\varepsilon} m$, for some 
constant $\varepsilon>0$.
We must also have that $m_r = m /\log^{2-r\varepsilon} m \leq m$, which implies that $r \leq2/\varepsilon$,
when $m$ is at least 3. When $\varepsilon$ is a fixed constant so is $r$. For example
when $\varepsilon = 1/4$ we get that $r$ is at most $2/(1/4)=8$.
Hence the maximum number of edges per static set follows a geometric progression.
%For instance, for $\varepsilon = 0.25$, we get $r = 8$ as $m\rightarrow\infty$ given that the condition $m\leq \sum_{i=0}^r m/\log^{2 - i\varepsilon} m$ must be satisfied.
Whenever we reach the maximum for a given set $E_i$, we find a set $E_j$, with $i < j \leq r$ such that $\sum_{\ell=0}^{j} m_\ell \leq m/\log^{2 - j\varepsilon} m$ and (re)build $E_j$ with all edges in it and in the previous sets, and reset all previous sets.
We detail this process below.

\SubSection{Space}\label{sec:static-to-dynamic:sec:space}
Let us analyse the required space to represent the data structure.
The set $E_0$ is represented in an uncompressed adjacency list coupled with a hash table to allow answering queries on edge existence in constant time.
If we use also a hash table to store the adjacency lists, then we need $\mathcal{O}(m_0\log m_0 + m_0\log n)$ bits, where $m_0\leq m/\log^{2} m$ is the number of edges in $E_0$.
Each set $E_i$, for $1\leq i\leq r$, is represented in a static $k^2$-tree and it requires $k^2 m_i \left(\log_{k^2}(n^2/m_i) + \mathcal{O}(1)\right)$ bits~\cite{Brisaboa2014}, where $m_i\leq m/\log^{2 - i\varepsilon} m$.
Hence, overall, the space required is
\begin{equation}\label{eq:overall_space}
\mathcal{O}(m_0\log m_0 + m_0\log n) + \sum_{i=1}^{r} k^2 m_i \left(\log_{k^2}(n^2/m_i) + \mathcal{O}(1)\right)
\end{equation}
bits. The first term in Equation~\ref{eq:overall_space} can be written as
$\mathcal{O}\left((m/\log^{2} m) (\log m + \log n)\right) = \mathcal{O}\left(m/\log n\right)$.
To bound the second term we essentially need to sum a geometric sequence, we will assume $m_i =  m /\log^{2-i\varepsilon}$ as this is the case
that requires more space. First let us sum the $m_i$ values,
\begin{equation}\label{eq:dynamic_space2}
	\sum_{i=1}^r m_i = m_1 \sum_{i=1}^r (\log^\varepsilon m)^{i-1} = m_1 \frac{(\log^{r\varepsilon}m)-1}{(\log^\varepsilon m) - 1}.
\end{equation}
Notice that as $m$ grows the logarithms dominate the values in the fraction which therefore approximates $(\log^{r\varepsilon}m)/\log^\varepsilon m$.
This expression can be further upper bounded by $(\log^{2}m)/\log^\varepsilon m$, because of the relation between $r$ and $\varepsilon$.
Hence the overall bound is the relation $\sum_{i=1}^r m_i \leq m_1 (\log^{2}m)/\log^\varepsilon m = m$

Now for the complete formula we obtain an upper bound by noticing that $m_i \geq m_0$ for all
$i$. The deduction is the following:
\begin{align*}
	 & \sum_{i=1}^{r} k^2 m_i \left(\log_{k^2}(n^2/m_i) + \mathcal{O}(1)\right) 
	   \leq  \sum_{i=1}^{r} k^2 m_i \left(\log_{k^2}(n^2/m_0) + \mathcal{O}(1)\right) \\
	 & = k^2 \left(\log_{k^2}(n^2/m_0) \left(\sum_{i=1}^{r} m_i \right)+ \mathcal{O}(r)\right) 
	   \leq  k^2 \left(m \log_{k^2}(n^2/m_0) + \mathcal{O}(1/\varepsilon)\right) \\
	 & \leq k^2 \left(m \left((2 \log \log n) + \log_{k^2}(n^2/m)\right) + \mathcal{O}(1/\varepsilon)\right).
\end{align*}

A tighter bound can be obtained by noticing that the largest terms in the sum are the last ones. Hence
essentially $m_r$ takes the role of $m_0$ in the previous expression, yielding an $\varepsilon \log \log n$ term 
instead of a $2 \log \log n$ term, which is expected to be reasonably small.

Therefore, the overall space in bits is bounded by
%\begin{equation}\label{eq:overall_space_2}
%\mathcal{O}\left(\frac{m}{\log^{2} m}(\log m + \log n)\right) + k^2 m \log_{k^2} \left(c \frac{n^2}{m} \right) \left(1 + o(1)\right).
	$k^2 m \left( \log_k(n^2/m) + 2 \log \log n \right) + \mathcal{O}(k^2/\varepsilon) + o(m)$.
%\end{equation}
% Moreover, for $m\geq n$, the space required is bounded by
% \begin{equation}\label{eq:overall_space_3}
% k^2 m \log_{k^2} \left(c \frac{n^2}{m} \right) \left(1 + o(1)\right).
% \end{equation}

\SubSection{Insertion, deletion and queries}
We rely on efficient set operations over $k^2$-trees~\cite{quijada2019set}.
Given $C$ and $C'$ represented as two $k^2$-trees, we are able to compute $k^2$-trees representing $C\cup C'$, $C\cap C'$ and $C\setminus C'$ in linear time on the size $|C|$ and $|C'|$ of the representations.
Moreover these operations are done without uncompressing $C$ and $C'$, with only some negligible extra space being used.

Insertion works as follows. Given a new edge $(u,v)$,
\begin{enumerate}
\item If $|E_0| < m_0$, then just add $(u,v)$ to $E_0$ and we are done.
\item Otherwise, build a $k^2$-tree for $E_0$,
find $0<j\leq r$ such that $\sum_{i=0}^j m_i \leq m_j$,
and rebuild $E_j$ with all edges in $E_0,\ldots,E_j$ by successive unions of $k^2$-trees.
\end{enumerate}
If $|E_0| < m_0$, then insertion takes constant time since we are relying on an adjacency list
coupled with a hash table to maintain adjacencies, as described before.
Otherwise, we need to build a $k^2$-tree for $E_0$ and find some $E_j$ to accommodate all
previous collections $E_i$, for $0\leq i \leq j$.
Note that the construction of the $k^2$-tree for $E_0$ takes $\mathcal{O}(m_0\log_k n)$
time~\cite{Brisaboa2014}, and the pairwise union of at most $j$ $k^2$-trees representing 
collections $E_0\ldots E_{j-1}$ takes
$\mathcal{O}(m_j\log_k n)$ time, using only the required space to store a $k^2$-tree representing $E_j$.
The amortized analysis of the insertion cost follows the argument presented by 
Munro {\em et al.}~\cite{DBLP:conf/pods/MunroNV15} for the general case.
Either $E_j$ is new and $m$ has at least doubled, in which case the amortized cost is
$\mathcal{O}(\log_k n)$ per edge insertion, or $E_j$ is not new and we are adding to it all
edges in collections $E_0,\ldots,E_{j-1}$. In this last case the building cost can be imputed to
the new edges added to $E_j$, which are at least $m_{j-1} = m_j / log^\varepsilon m$.
Therefore, the amortized cost of inserting an edge in $E_j$ is 
$\mathcal{O}(\log_k n \log^\varepsilon m)$ and, since
each link can be moved once to each $E_j$, with $0<j\leq r = \lfloor 2/\varepsilon \rfloor$, the
amortized cost of inserting an edge is $\mathcal{O}(\log_k n \log^\varepsilon m (1/\varepsilon))$.
And this is then the overall amortized cost of inserting an edge.

Deletion works as follows. Given an edge $(u,v)\in E$,
\begin{enumerate}
\item If $(u,v)\in E_0$, then just remove it and we are done.
\item Otherwise, find $0<j\leq r$ such that $(u,v)\in E_j$ and, if there is such $j$, set the corresponding bit to zero in $E_j$ $k^2$-tree.
\item Update the number $m'$ of deleted edges.
\item If $m' > m / \log\log m$, rebuild $\mathcal{C}$.
\end{enumerate}
Deleting and edge in $E_0$ takes constant time.
Checking and deleting an edge in our collections takes
$\mathcal{O}((\log_k n)/\varepsilon)$,
since checking if an edge exists in a given $k^2$-tree takes $\mathcal{O}(\log_k n)$~\cite{Brisaboa2014}, and
we might have to look in each collection $E_i$, 
with $0<i\leq r = \lceil 2/\varepsilon \rceil$.
Once an edge is found, marking it for deletion takes constant time.
The full rebuild after $m / \log\log m$ edges are deleted costs $\mathcal{O}(m\log_k n)$,
\ie, it has an amortized cost of $\mathcal{O}(\log_k n \log\log m)$ per deleted edge.
Overall deleting an edge has then an amortized cost of $\mathcal{O}((\log_k n)/\varepsilon + \log_k n \log\log m)$.

Querying works just as in $k^2$-trees with the difference that we need to query all sets in the collection.
Therefore, the querying cost increases by a factor of $\mathcal{O}(1/\varepsilon)$.

\SubSection{Comparison with other constructions}\label{sec:static-to-dynamic:sec:comparison}
Given a graph $G$, for a fixed $\varepsilon$, the presented data structure uses essentially the same space as 
a static $k^2$-tree, and it supports insertions and deletions
in $\mathcal{O}(\log_k n \log^\varepsilon m)$
and $\mathcal{O}(\log_k n \log\log m)$ time, respectively.
The implementation of dynamic $k^2$-trees using dynamic bit vectors~\cite{DBLP:journals/is/BrisaboaCBN17}
requires a small space overhead, and it supports insertions and deletions in
$\mathcal{O}(\log_k n \log n)$ time.
Hence, since $m$ is $\mathcal{O}(n^2)$, it has a slowdown by a factor of $o(\log n / \log\log n)$ with
respect to the proposed data structure.

Edge queries over the proposed data structure take the same time as in static $k^2$-trees.
Although dynamic $k^2$-trees using dynamic bit vectors~\cite{DBLP:journals/is/BrisaboaCBN17} 
work similarly to static $k^2$-trees -- in practice they replace static bit vectors for dynamic ones -- they
suffer a slowdown by a factor of $\Omega(\log n / \log\log n)$~\cite[Chapter 12]{navarro2016compact}.

We compare also with a new representation, $k^2$-tries, proposed recently~\cite{DBLP:conf/spire/ArroyueloBGN19}.
This data structure uses $\mathcal{O}(m\log(n^2/m) + m \log k)$ bits, and it supports edge queries and updates in
$O(\log_k n)$ amortized time. The implementation provided by $k^2$-tries authors supports only edge additions
and queries, with slightly worse time complexities.

\Section{Experimental analysis}\label{sec:experimental-analysis}
We compare the dynamic $k^2$-tree implementation proposed in this paper, henceforth named
\texttt{sdk2tree}, with the dynamic implementation \texttt{dk2tree} based on
dynamic bit vectors~\cite{DBLP:journals/is/BrisaboaCBN17}, a
static implementation \texttt{k2tree}~\cite{Brisaboa2014}, and also
with two versions of $k^2$-tries, \texttt{k2trie\{1,2\}}, that differ only on the parametrization (trading compression for speed)~\cite{DBLP:conf/spire/ArroyueloBGN19}.
All other implementations were provided by their authors, and all code is available at \url{https://github.com/aplf/sdk2tree}.

All tested implementations are written in $C$ and compiled with \texttt{gcc
6.3.0 2017\-05-16} using the \texttt{-O3} optimization flag.  Experiments were
performed on an SMP machine with 256GB of RAM and four Intel(R) Xeon(R) CPU
E7-4830 @ 2.13GHz, each one with 512KB in L1 cache, 2MB in L2 cache, 24MB in L3
cache and eight cores, 64 threads in total. All implementations are
single-threaded.

We implemented a common interface to test each implementation. All dynamic data structures
\texttt{dk2tree}, \texttt{sdk2tree} and \texttt{k2trie\{1,2\}}
are initialized empty. The static \texttt{k2tree} is initialized by reading the whole
graph from secondary storage. 
Once initialized, the interface starts
a main loop which reads instructions from \texttt{stdin} representing
all supported edge operations, with additions and deletions not available in \texttt{k2tree},
and \texttt{k2trie\{1,2\}} supporting only edge additions and queries.

\SubSection{Datasets and methodology}\label{sec:experimental-analysis:sec:datasets}
We used both real and synthetic datasets.
In Table~\ref{table:datasets} we identify the datasets and their properties.
For each dataset, we present its vertex and edge counts written as $|V|$ and $|E|$, respectively, and the total disk space used by each implementation.
%We have graphs with vertex and edge counts of different orders of magnitude.

Real-world graphs were obtained from the Laboratory of Web Algorithmics\footnote{\url{http://law.di.unimi.it/datasets.php}}~\cite{BoVWFI,BRSLLP}.
Synthetic datasets were generated from the partial duplication
model~\cite{chung2003duplication}.  Although the abstraction of real networks
captured by the partial duplication model, and other generalizations, is rather
simple, the global statistical properties of, for instance, biological networks
and their topologies can be well represented by this kind of
model~\cite{bhan2002duplication}.
% For each number of vertices, we generated 10 
We generated random graphs with selection probability $p=0.5$, which is within
the range of interesting selection probabilities~\cite{chung2003duplication}.
The number of edges for those graphs is approximately $25$ times the number of
vertices.

%The \texttt{k2tree} version builds its representation and stores it on disk from an edge binary adjacency list format.
%%The \texttt{k2tree} version builds the $k^2$-tree representation and stores it in disk beforehand, it was necessary to perform conversions.
%The other two versions read operations from standard input in a tab-separated format.
%To obtain the datasets in all required input formats, we implemented in \texttt{Java} the following converters: \textit{a)} from our tab-separated format into the binary adjacency list for the \texttt{k2tree} version; \textit{b)} from the downloaded Web graph compressed format to the same binary adjacency list; \textit{c)} from the Web graph compressed format to our tab-separated format.

\begin{table}
% tabluarx: the last column needs an \arraybackslash to be able to center:
% https://tex.stackexchange.com/questions/148912/why-cannot-the-word-in-the-last-column-be-centered
\caption{The first four datasets were synthetically generated using a duplication model.
The last five datasets are real-world Web graphs made available by the Laboratory for Web Algorithmics (LAW)~\cite{BoVWFI,BRSLLP}
(dataset \texttt{uk-2007-05} is actually \texttt{uk-2007-05-100000} in the LAW website).}
\label{table:datasets}
{\small
\begin{tabularx}{\textwidth}{
>{\centering}m{2.80cm}  | 
>{\centering}m{0.90cm}  | 
>{\centering}m{1.15cm}  | 
>{\centering}m{1.30cm}  | 
>{\centering}m{1.45cm}  | 
>{\centering}m{1.45cm}  | 
>{\centering}m{1.45cm}  | 
>{\centering\arraybackslash}m{1.45cm} 
}
\textbf{Dataset} & $|V|$\\{\footnotesize (M)} & $|E|$\\{\footnotesize (M)} & \texttt{k2tree} {\footnotesize (MB)} & \texttt{dk2tree} {\footnotesize (MB)} & \texttt{sdk2tree} {\footnotesize (MB)} & \texttt{k2trie1} {\footnotesize (MB)} & \texttt{k2trie2} {\footnotesize (MB)} \\ \hline
\end{tabularx}

% NOTE: we use \vskip to bring the two `tabularx' environments together:
% https://tex.stackexchange.com/questions/30062/vspace-vs-vskip
\vskip -0.03cm

% sinunitx: vertical alignment of the last column - problems:
% https://tex.stackexchange.com/questions/316453/last-column-problem-with-csvsimple-and-siunitx
% NOTE: the solution for the last column was to add a dummy `p' column.
\begin{tabularx}{\textwidth}{
m{2.80cm} | 
S[table-format = 2.3, table-number-alignment = center, table-column-width = 0.90cm] |
S[table-format = 3.3, table-number-alignment = center, table-column-width = 1.15cm] |
S[table-format = 3.3, table-number-alignment = center, table-column-width = 1.30cm] |
S[table-format = 3.3, table-number-alignment = center, table-column-width = 1.45cm] |
S[table-format = 3.3, table-number-alignment = center, table-column-width = 1.45cm] |
S[table-format = 3.3, table-number-alignment = center, table-column-width = 1.45cm] |
S[table-format = 3.3, table-number-alignment = center, table-column-width = 1.45cm]
}

\texttt{dm50K}             &  0.05 &   1.11 &   2.80 &   3.13 &   2.82 &   5.72 &  39.61 \\ %  2.789
\texttt{dm100K}            &  0.10 &   2.59 &   7.01 &   7.82 &   7.04 &  14.63 &  79.65 \\ %  6.997
\texttt{dm500K}            &  0.50 &  11.98 &  39.80 &  44.05 &  39.95 &  82.71 & 268.34 \\ % 39.674
\texttt{dm1M}              &  1.00 &  27.42 &  96.35 & 111.82 &  96.39 & 192.12 & 434.54 \\ \hline %101.018

\texttt{uk-2007-05}        &  0.10 &   3.05 &   1.08 &   1.23 &   1.15 &   2.05 &   4.04 \\ %  1.084
\texttt{in-2004}           &  1.38 &  16.92 &   6.04 &   6.85 &   6.32 &   7.86 &  14.05 \\ %  6.015
\texttt{uk-2014-host}      &  4.77 &  50.83 &  57.41 &  63.94 &  58.05 &  79.22 & 132.61 \\ % 57.375
\texttt{indochina-2004}    &  7.42 & 194.11 &  56.90 &  64.46 &  59.87 &  66.72 & 113.53 \\ % 56.799
\texttt{eu-2015-host}      & 11.26 & 386.92 & 258.87 & 288.66 & 263.27 & 323.68 & 537.04    %257.340
\end{tabularx}
}
\end{table}

We consider four major operations: edge additions,
removals, querying/checking and vertex neighborhood listing.
Elapsed time was measured using the \texttt{clock()}
function\footnote{\url{http://man7.org/linux/man-pages/man3/clock.3.html}}.
%
%\textbf{Addition.}
Although the \texttt{k2tree} implementation does not support additions, we
included it in the comparison. For that we build a \texttt{k2tree} for each dataset and we
divided the time it took by the number of edges, obtaining then the average
time for edge addition. This allowed us to evaluate the overhead introduced by
dynamic data structures.
%%The addition operation is evaluated solely between \texttt{sdk2tree} and
%%\texttt{dk2tree}, as it makes no sense to consider the addition in the case of
%%the \texttt{k2tree} implementation.  
%
%To test the additions in the other implementations, after the initialization routines, we pipe into each
%implementation a sequence of edge addition instructions (e.g., \texttt{``a 1 2''} is
%the instruction to add edge \textit{(1, 2)}) from a tab-separated edge list (one edge
%per line). We then measure how much time elapsed (after initialization) from
%the moment the first edge addition instruction was received until the time the last
%edge addition is performed (
%
%\textbf{Removal.} As with additions,
The removal operation is compared only between \texttt{sdk2tree} and
\texttt{dk2tree}.
%The removal operation is compared against the performance of other operations
%for the \texttt{sdk2tree} implementation, as the \texttt{dk2tree}
%implementation~\cite{DBLP:journals/is/BrisaboaCBN17} did not function properly
%when removals were attempted.
This operation was evaluated by adding all edges and removing a sample of 50\% of them.
%To prepare the evaluation of this operation, for each evaluated graph, we take
%an offline sample of 50\% of its original edges and store the sample in a
%list of edge deletion instructions. %(one edge removal per line).
%After adding all edges, we process then these instructions.
%(e.g., \texttt{``d 1 2''} is an instructions to remove
%edge \textit{(1, 2)}). The removal operation time is measured likewise -- from
%the moment the first edge removal instruction is received until the processing
%of the last edge removal instruction finishes.
%
%\textbf{Listing.}
All three \texttt{*k2tree} implementations were directly compared for the listing operation.
After adding all edges, we evaluated this operation by asking for the neighborhoods of a sample of 50\% of the vertices.
We measure for each implementation the average time per individual operation,
the maximum resident
set size (memory peak was obtained with \texttt{GNU time}\footnote{\url{https://www.gnu.org/software/time/}}), and the disk space
taken by data structures serialization.
%When comparing peak
%resident memory size, for the \texttt{k2tree} implementation, we show both the
%memory reported by \texttt{GNU time} and also the size that the $k^2$-tree
%serialization took in secondary storage.

\begin{figure}
    \centering
		\includegraphics[width=.495\textwidth]{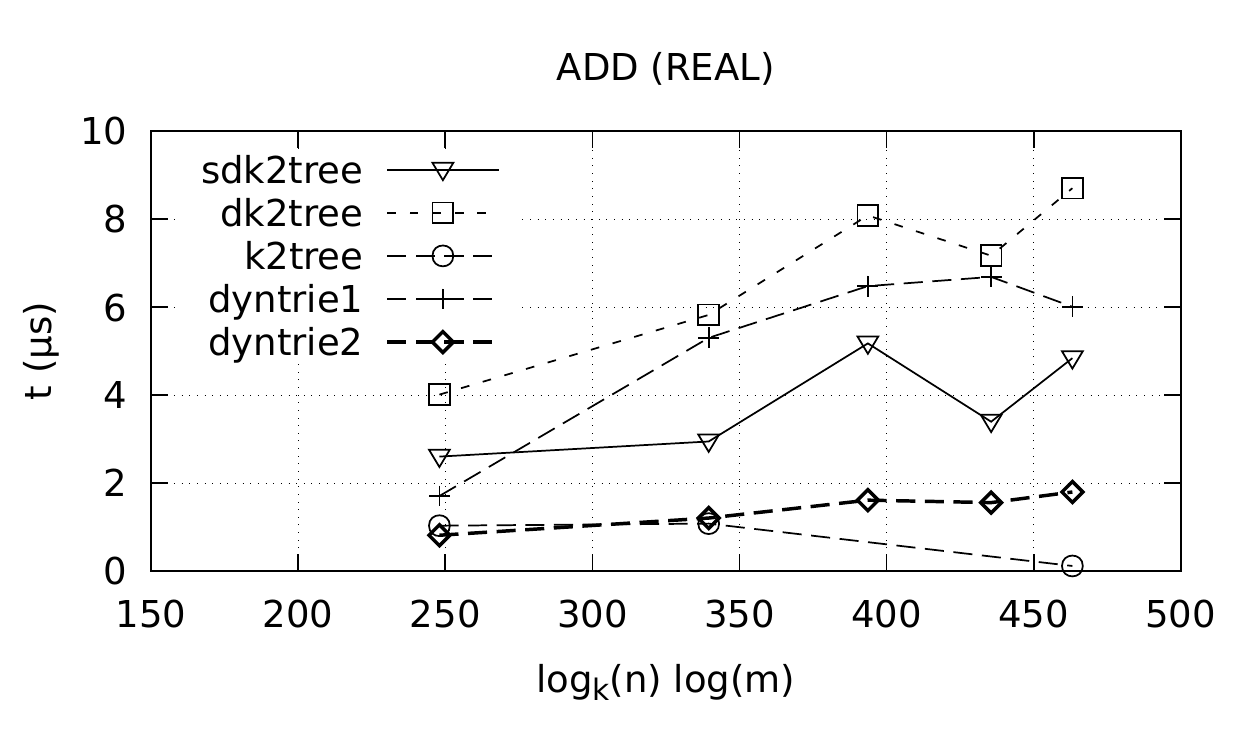}
		\includegraphics[width=.495\textwidth]{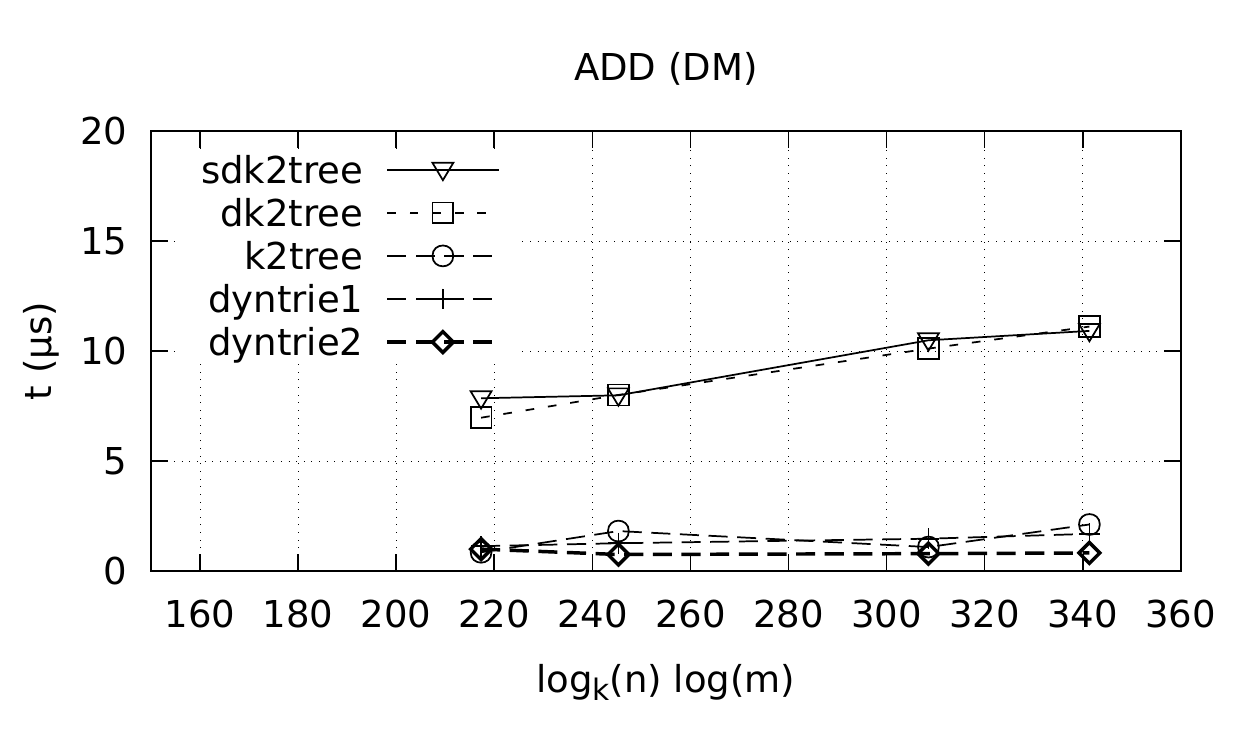}
		\caption{Average time taken for adding an edge in real Web graphs
    and in synthetic graphs (generated from a duplication model), respectively.}
    \label{fig:A-time}
\end{figure}

\begin{figure}
    \centering
		\includegraphics[width=.495\textwidth]{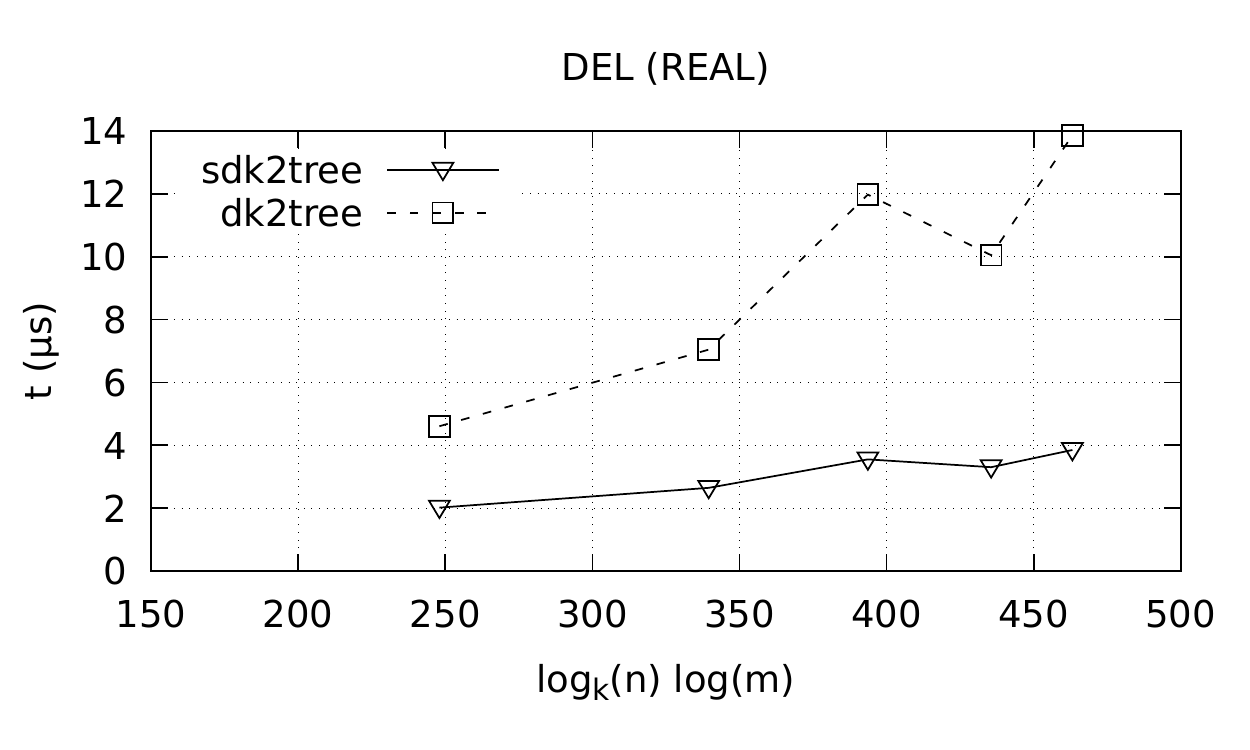}
		\includegraphics[width=.495\textwidth]{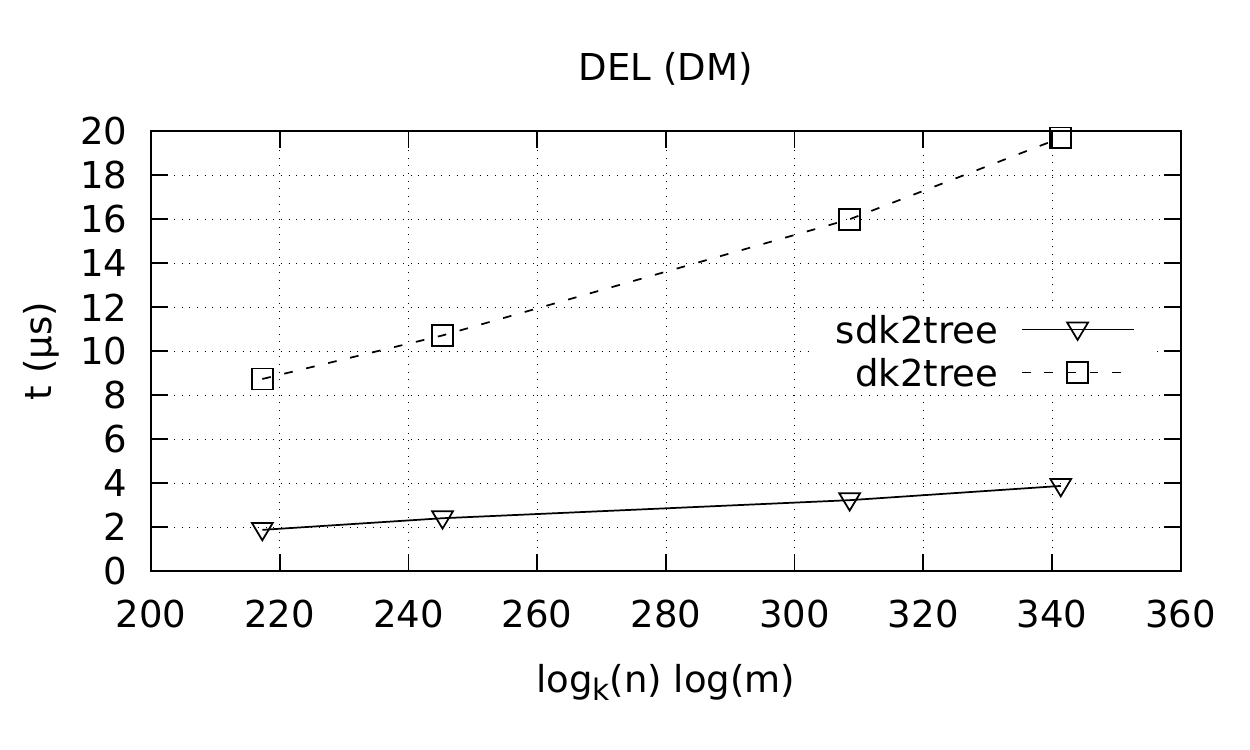}
		\caption{Average time taken for deleting an edge in real Web graphs
    and in synthetic graphs (generated from a duplication model), respectively.}
    \label{fig:AD-time}
\end{figure}

\SubSection{Cost analysis}\label{sec:experimental-analysis:sec:performance}
Let us analyse the cost of each operation over the different datasets and for
the different implementations. Figure~\ref{fig:A-time} shows the average
running time for adding an edge.  As mentioned before, we
included \texttt{k2tree} in this comparison to observe what is the slowdown introduced by dynamic data structures. As
expected, dynamic implementations take in general more time per add
operation than \texttt{k2tree}. We can observe that \texttt{k2trie\{1,2\}} and \texttt{sdk2tree} are sometimes slightly faster
than \texttt{k2tree}. The case of \texttt{sdk2tree} may be explained by the sparsity of the internal $k^2$-trees.
As expected also from the theoretical
analysis, the add operation on \texttt{sdk2tree} is faster than on \texttt{dk2tree}, in particular for
real Web graphs.
Figure~\ref{fig:AD-time} shows the average running time for removing an edge.
Across all datasets, \texttt{sdk2tree} was
consistently faster than \texttt{dk2tree}.
We note that costs seem to correlate well with the predicted bounds.

\begin{figure}
    \centering
		\includegraphics[width=.495\textwidth]{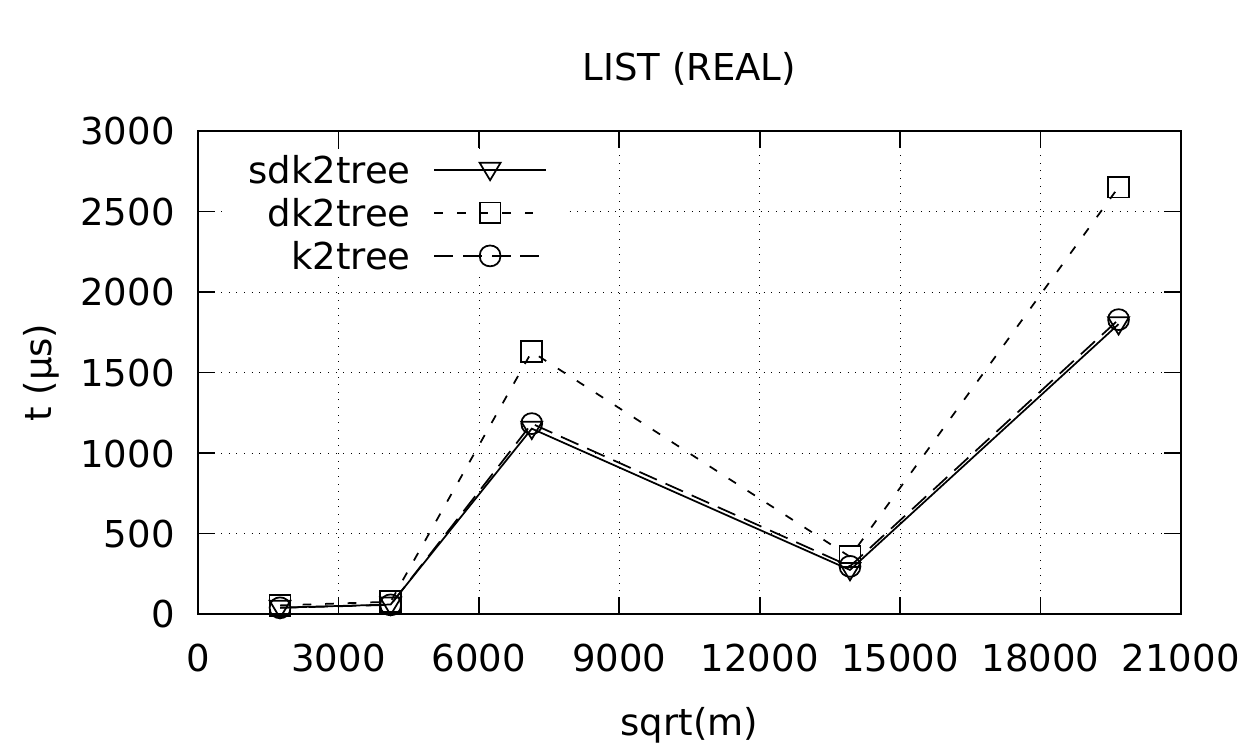}
		\includegraphics[width=.495\textwidth]{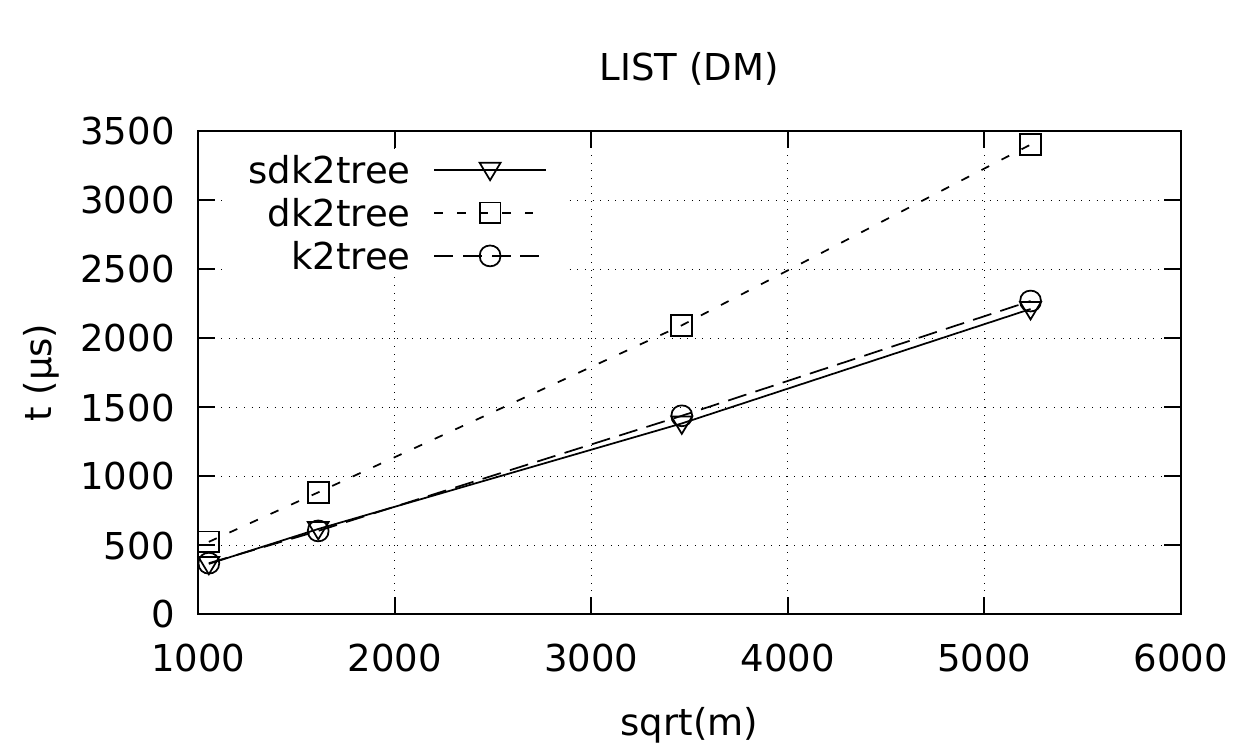}
		\caption{Average time taken for listing neighbors of random vertices in real Web graphs
    and in synthetic graphs (generated from a duplication model), respectively.}
    \label{fig:AL-time}
\end{figure}

\begin{figure}
    \centering
		\includegraphics[width=.495\textwidth]{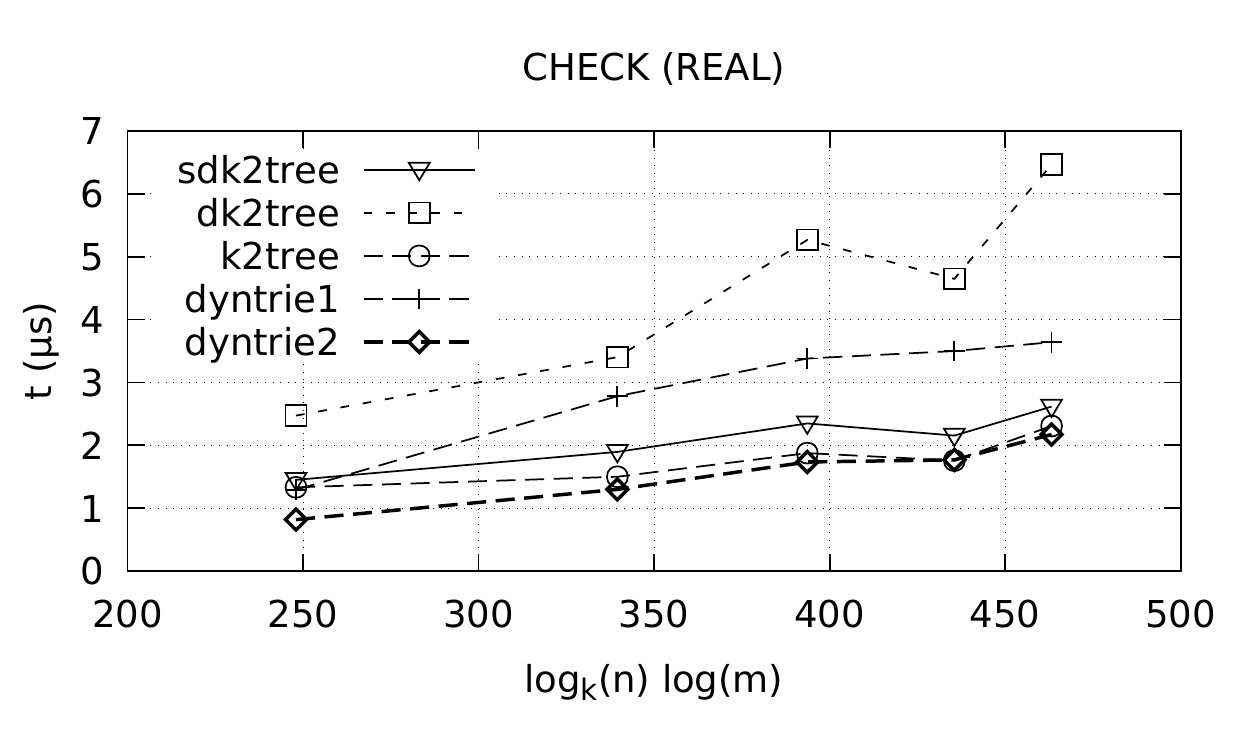}
		\includegraphics[width=.495\textwidth]{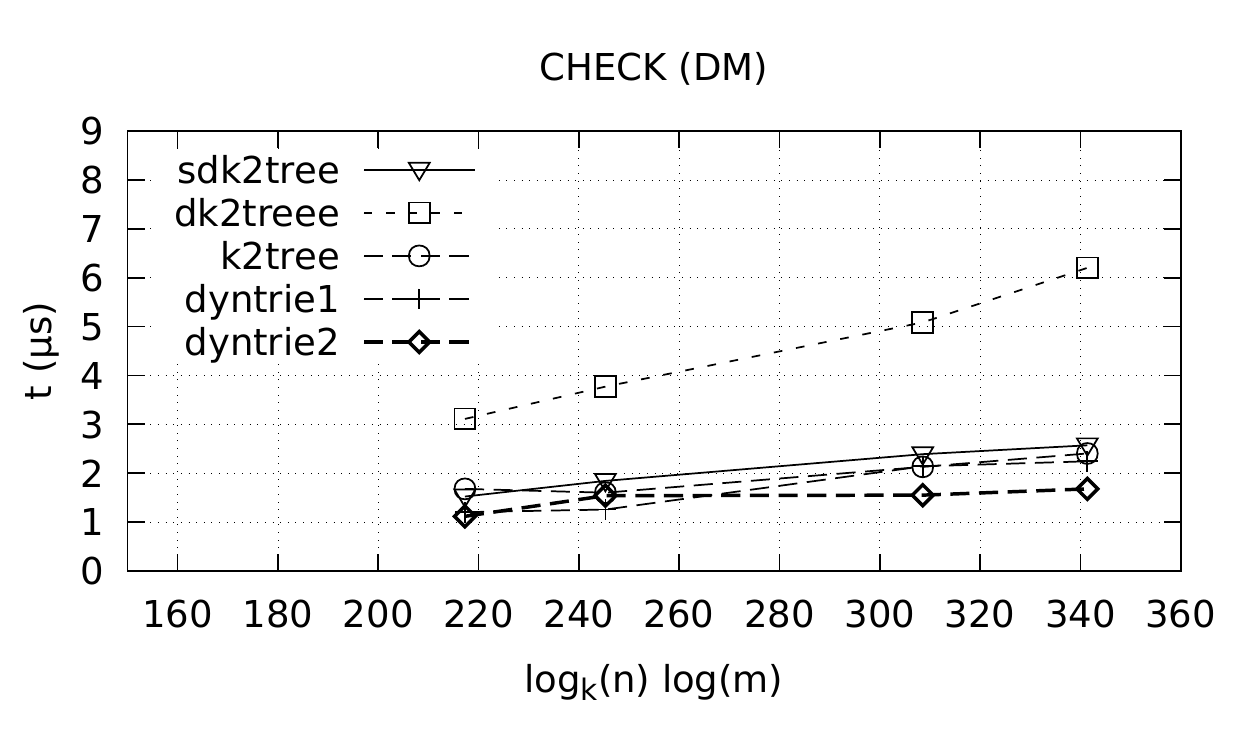}
		\caption{Average time taken for querying links in real Web graphs
    and in synthetic graphs (generated from a duplication model), respectively.}
    \label{fig:AC-time}
\end{figure}

Figures~\ref{fig:AL-time} and~\ref{fig:AC-time} show the average running time for listing vertex
neighborhoods and querying/checking edges. Across all datasets, \texttt{sdk2tree} was faster than
\texttt{dk2tree} and on-par with \texttt{k2tree} and \texttt{k2trie\{1,2\}}.
In the case of listing, we are plotting against $\mathcal{O}(\sqrt{m})$,
the bound on the cost of listing vertex neighborhoods with \texttt{k2tree}~\cite{Brisaboa2014}.
This bound is valid also for
\texttt{sdk2tree} and \texttt{dk2tree} as discussed previously in the theoretical analysis.

\begin{figure}
    \centering
		\includegraphics[width=.495\textwidth]{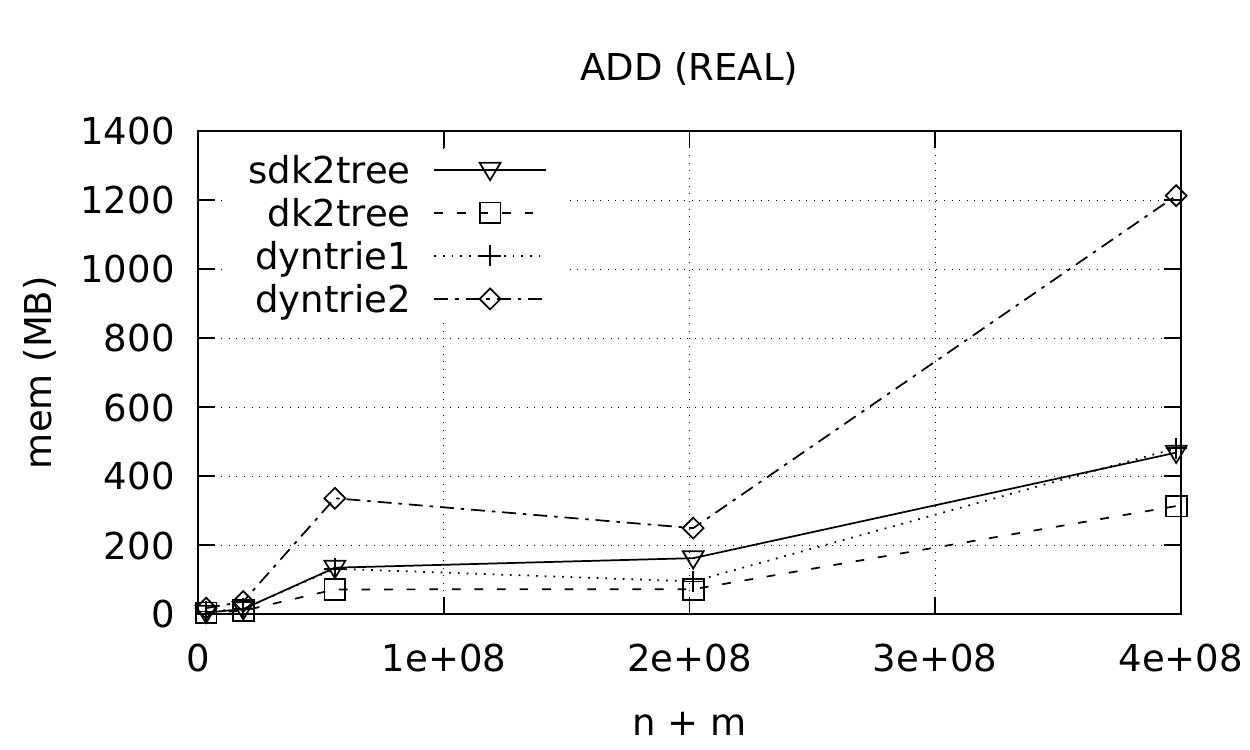}
		\includegraphics[width=.495\textwidth]{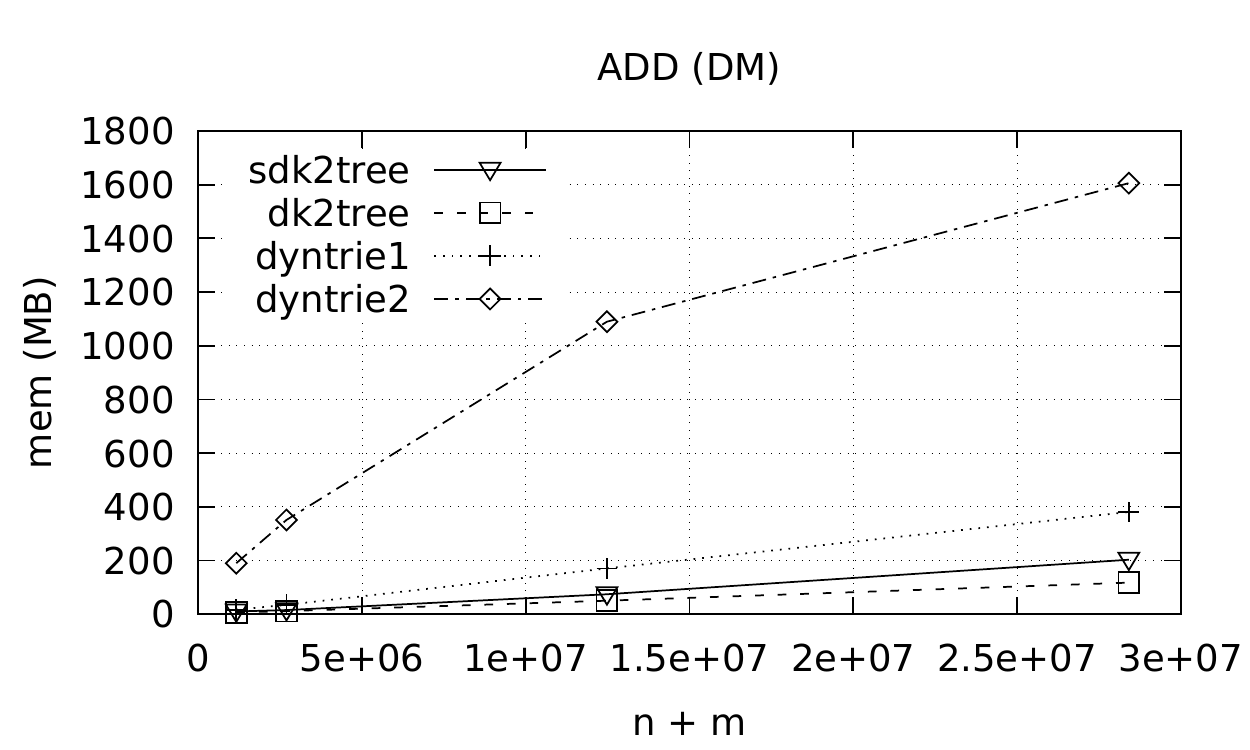}
		\caption{Max resident memory while adding edges in real Web graphs
    and in synthetic graphs (generated from a duplication model), respectively.}
    \label{fig:A-memory}
\end{figure}

Let us now analyse how much memory is used by each implementation.
In this analysis we will consider resident memory while we are performing
operations. For the space that each data structure takes once serialized on
secondary memory, we refer the reader to Table~\ref{table:datasets}.
Figure~\ref{fig:A-memory} shows the max resident memory while adding edges in dynamic implementations.
We can observe that \texttt{sdk2tree}
requires more memory than \texttt{dk2tree}, although the growth rate is
similar. This can look unexpected given the theoretical bounds derived
previously, but we must recall that we are periodically merging together
static collections in the \texttt{sdk2tree} implementation. We will analyse
this in more detail below.

\begin{figure}
    \centering
		\includegraphics[width=.495\textwidth]{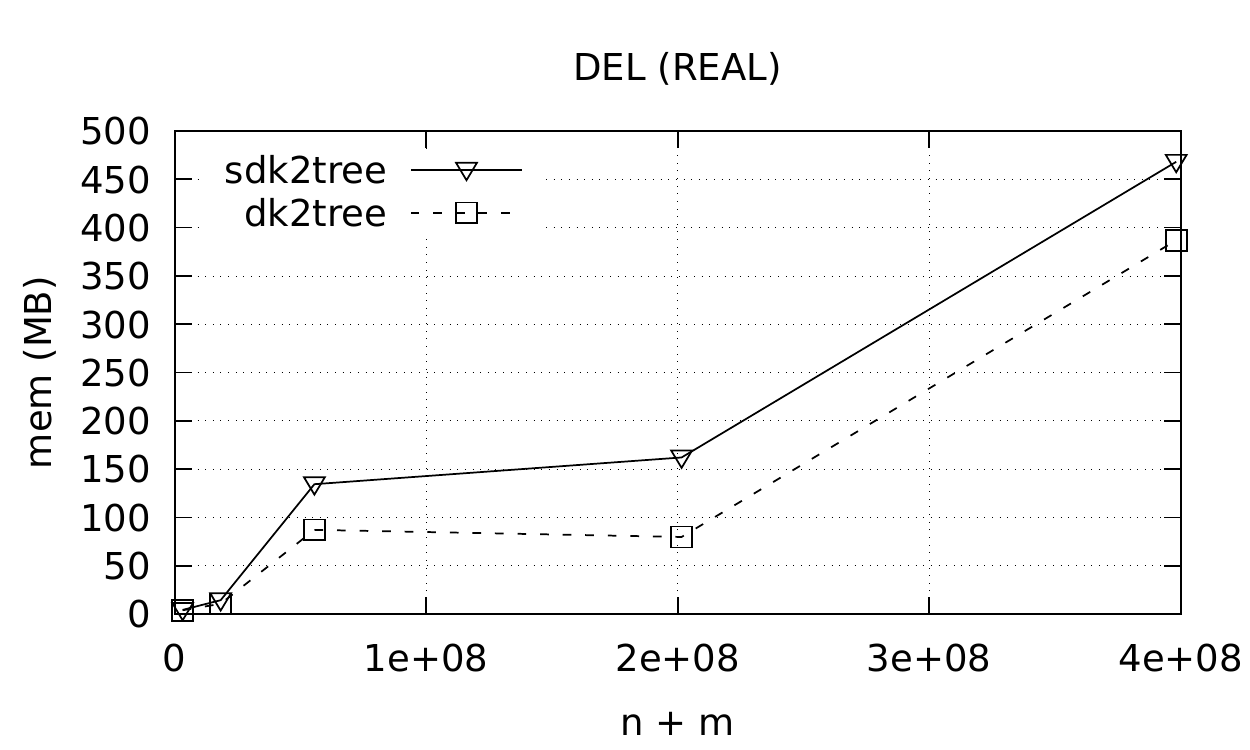}
		\includegraphics[width=.495\textwidth]{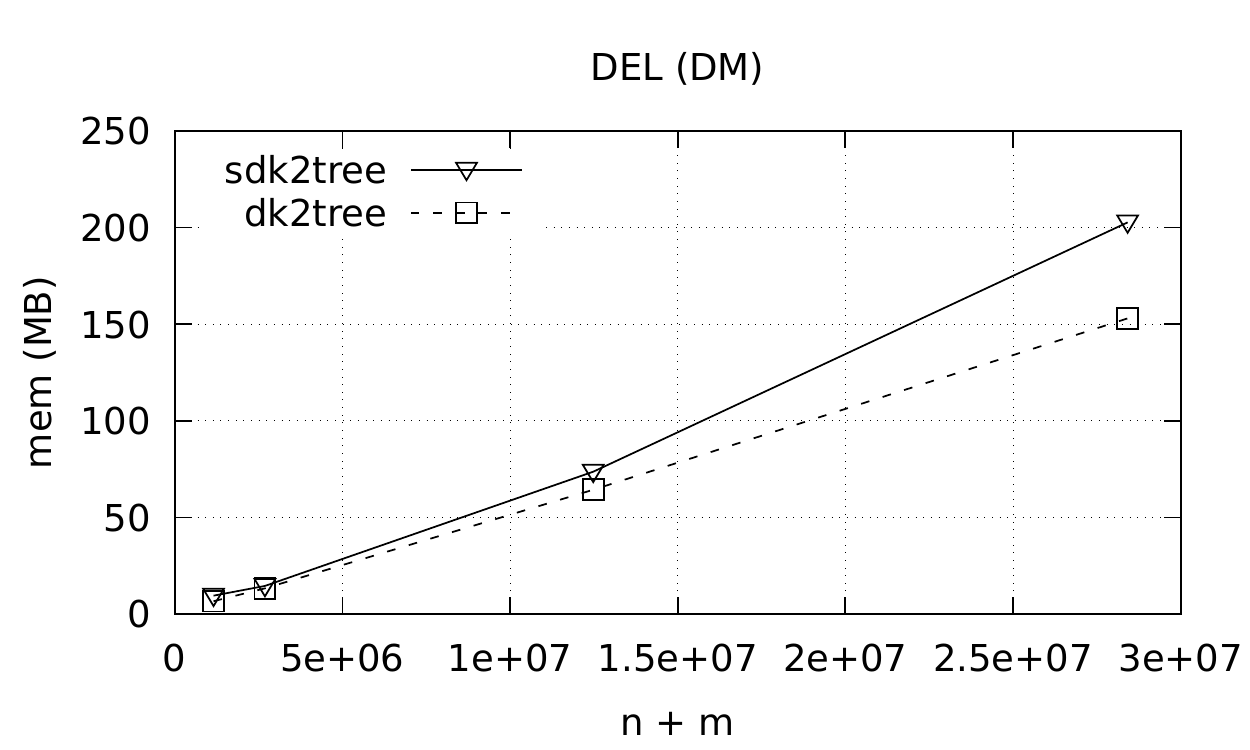}
		\caption{Max resident memory while deleting edges in real Web graphs
    and in synthetic graphs (generated from a duplication model), respectively.}
    \label{fig:AD-memory}
\end{figure}

Figure~\ref{fig:AD-memory} shows the max resident memory while removing edges.
Since we are adding all links before removing about 50\% of them, the memory
requirements for \texttt{sdk2tree} are exactly the same as in Figure~\ref{fig:A-memory}.
This also means that the removing operation  does not increase the space requirements
in this implementation. On the other hand, the memory requirements are now higher for
\texttt{dk2tree}, being more close to those of \texttt{sdk2tree}.

\begin{figure}
    \centering
		\includegraphics[width=.495\textwidth]{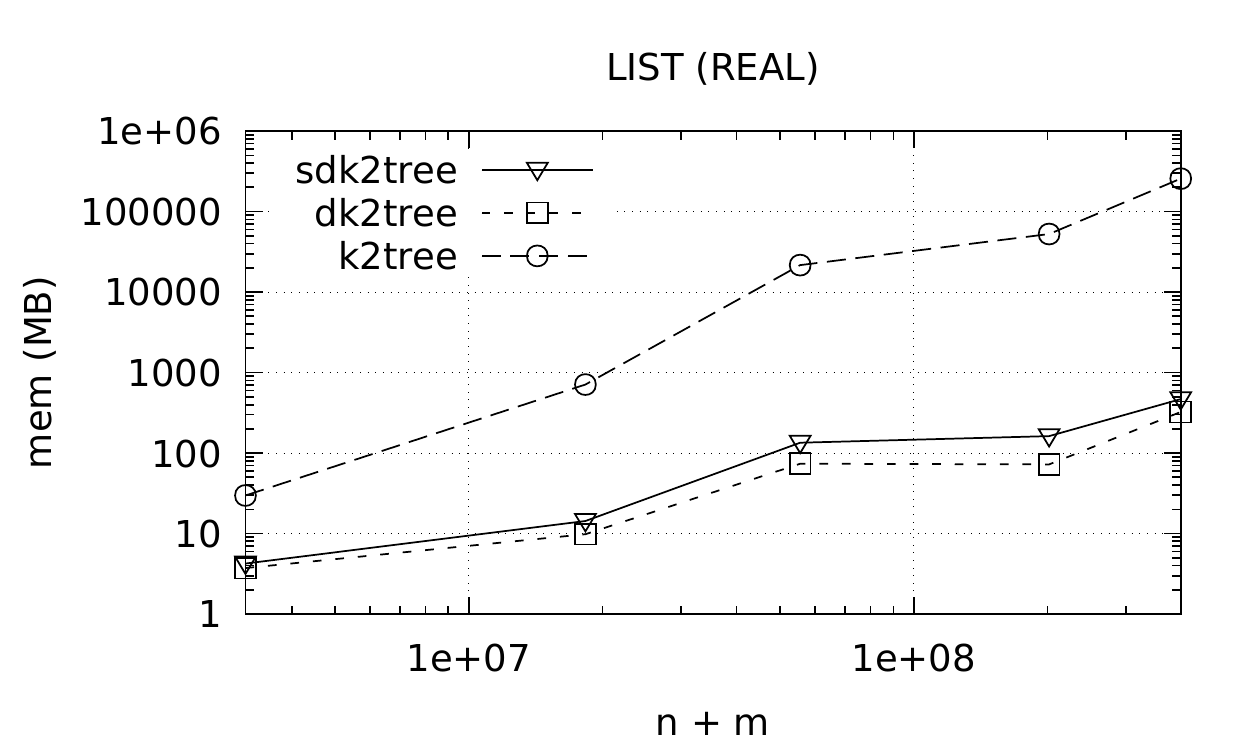}
		\includegraphics[width=.495\textwidth]{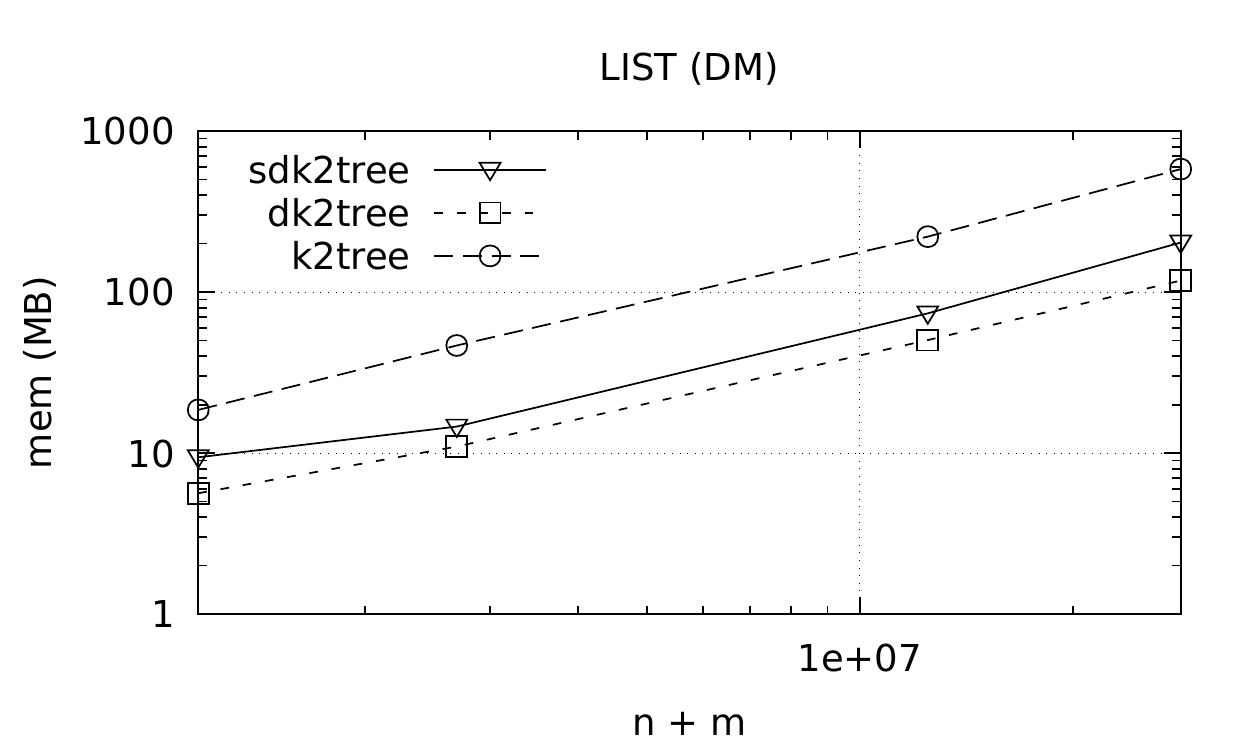}
		\caption{Max resident memory while listing neighbors of random vertices in real Web graphs
    and in synthetic graphs (generated from a duplication model), respectively.}
    \label{fig:AL-memory}
\end{figure}

Figure~\ref{fig:AL-memory} shows the max resident memory while adding edges and listing
vertex neighborhoods. Since we are adding all links as before, the memory requirements
for \texttt{sdk2tree} and \texttt{dk2tree} are identical to those observed in
Figures~\ref{fig:A-memory} and~\ref{fig:AD-memory}. We included now also the
\texttt{k2tree} in our analysis. Given that this last implementation requires
much more space for constructing the data structure, we had to use log scale
in Figure~\ref{fig:AL-memory}. We should note however that once constructed,
\texttt{k2tree} requires much less space as shown in Table~\ref{table:datasets}.
For instance, for the dataset \texttt{dm100K},
\texttt{k2tree} had a peak resident memory footprint of around 503.11 MB during
construction, while its $k^2$-tree structure stored on disk is around 7.01 MB.
It is nevertheless interesting to note that, although we are using the
exact same implementation of $k^2$-trees for representing the static collections
within our \texttt{sdk2tree} implementation, since we are merging those collections
without decompressing them as mentioned before, we do not observe such high memory
footprint while adding edges in \texttt{sdk2tree}.

%Figure~\ref{fig:AL-memory-log} also portrays for the \texttt{k2tree} version the contrast between the peak memory footprint (noted as \texttt{static (inc. full reconstruction from disk)}) against the size of the corresponding $k^2$-tree on disk (\texttt{static-on-disk}).
%Effectively, the \texttt{k2tree} implementation incurs a memory overhead when loading the graph and building the internal representation, such that it is orders of magnitude greater than the size on disk.
%For example, for dataset \texttt{100000} (0.1 M vertices) the \texttt{k2tree} version had a peak resident memory footprint of around 503.112 MB while its $k^2$-tree structure stored on disk is around 7.007 MB.
%Regarding deletions, Fig.~\ref{fig:AD-memory-log} shows the same memory footprint relation between \texttt{sdk2tree} and \texttt{dk2tree} that was shown in Fig.~\ref{fig:A-memory-log}.

%%%% dyn-upc operation log times.
%The \texttt{dk2tree} implementation shows similar proportions between addition time and neighborhood listing time.
%In Fig.~\ref{fig:dyn-upc-total-time-log} we see the log-scale total time for each operation type while Fig.~\ref{fig:dyn-upc-op-time-log} shows the log-scale time per operation.
%They show the same difference in magnitude for \texttt{dk2tree} as seen in \texttt{sdk2tree}.
%We reiterate that the removal operation is faster in \texttt{sdk2tree} than in \texttt{dk2tree} as previously shown in Fig.~\ref{fig:AD-ops}.
%This is important to note, as it explains how in Fig.~\ref{fig:dyn-upc-total-time-log} (\texttt{dk2tree}) the cumulative time taken by the addition operation is greater than the cumulative removal operation time, despite the average time of the removal operation being greater than the average time of the addition operation in Fig.~\ref{fig:dyn-upc-op-time-log}.
%Although the average time of the removal operation in \texttt{dk2tree} was lower than the average time of the addition, the amount of addition operations for each dataset surpassed (number of removals was sampled as 50\% of the number of additions) the amount of removals, resulting in this behavior.

\SubSection{Memory allocation analysis}\label{sec:experimental-analysis:sec:allocation_overhead}

\begin{figure}
    \centering
    \includegraphics[width=0.9\textwidth]{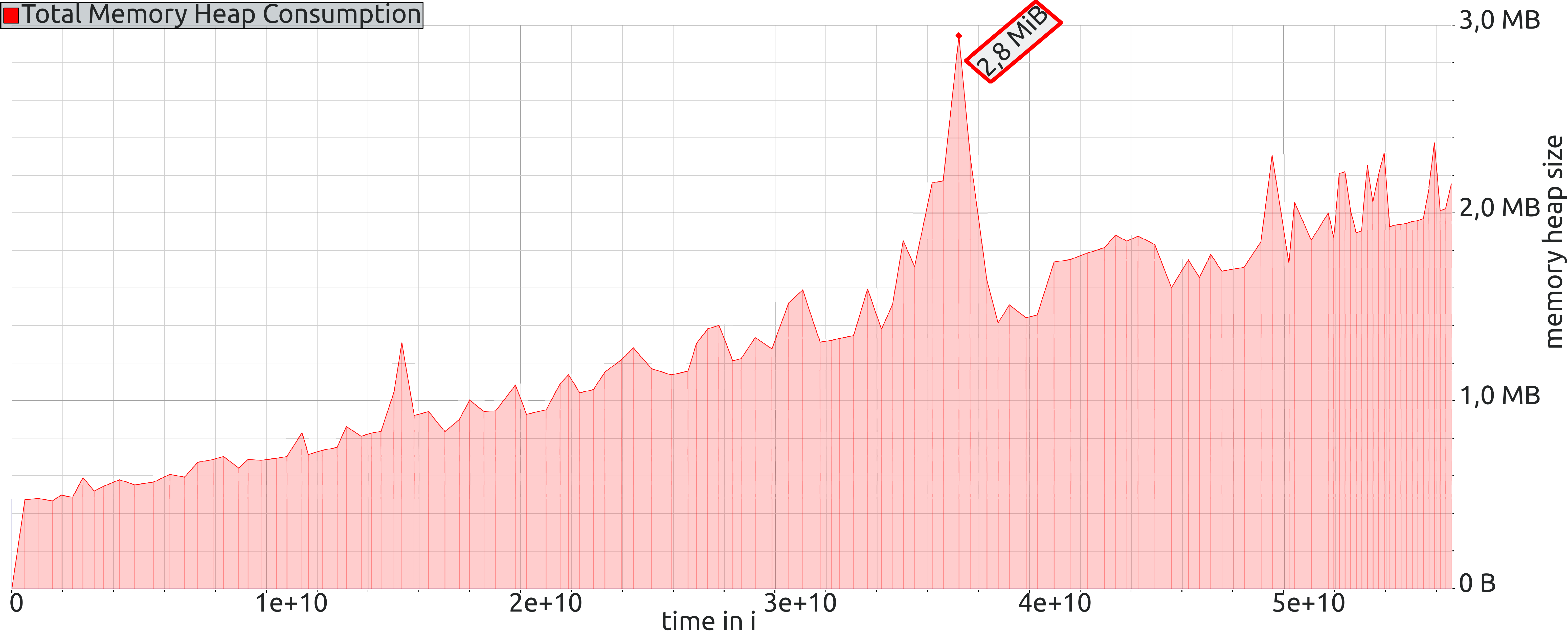}
    %\caption{\texttt{valgrind} heap allocation profile for datasets \texttt{dm1M} and \texttt{uk-2007-05}, respectively. The label \texttt{time in i} in the $x$ axis denotes the number of instructions
    \caption{\texttt{valgrind} heap allocation profile for dataset \texttt{uk-2007-05}. The label \texttt{time in i} in the $x$ axis denotes the number of instructions
executed.}
    \label{fig:valgrind_profile}
\end{figure}

Our implementation of the dynamic $k^2$-tree is based on the technique
presented in~\cite{DBLP:conf/pods/MunroNV15}, whose authors claim additional
space is necessary to perform a union of two collections (which would be decompressed before the union
operation taking place).
The implementation we present is able to perform the union operation without
decompressing the collections, effectively avoiding this pitfall.
We show for dataset %s \texttt{dm100K} and 
 \texttt{uk-2007-05}, in
Figure~\ref{fig:valgrind_profile}, a detailed analysis of heap memory usage.
The analysis was performed using \texttt{valgrind}, with parameters
\texttt{--tool=massif --max-snapshots=200 --detailed-freq=5}, and the
visualizations using the \texttt{massif-visuali\-zer}\footnote{\url{https://github.com/KDE/massif-visualizer}}.

It can be observed that during execution where edges are continuously added,
there are memory peaks associated with the union operation, increasing
temporarily the heap usage by at most a factor of 2. This explains also the
difference in maximum resident memory between \texttt{sdk2tree} and
\texttt{dk2tree} observed before in Figures~\ref{fig:A-memory}
and~\ref{fig:AD-memory}.

\Section{Final remarks}\label{sec:remarks}
We presented the \texttt{sdk2tree} implementation for representing dynamic
graphs, based on the $k^2$-tree graph representation and relying on a
collection of static $k^2$-trees. This makes \texttt{sdk2tree} a
semi-dynamic data structure.  Nevertheless, it supports edge additions and
removals with competitive performance, showing faster execution times than the
\texttt{dk2tree} implementation, a dynamic version of $k^2$-trees based on
dynamic bit vectors, and on par with $k^2$-tries with respect to additions and queries.

Implementations like those analysed in this paper, when implemented carefully,
are of crucial importance for the efficient analysis and storage of evolving
graphs, while drastically reducing the requirements of secondary storage
compared to traditional dynamic graph representations. Hence, as future work,
we envision further refinements to these data structures to achieve greater
efficiency, namely in what concerns listing vertex neighborhoods, in order
to produce usable libraries for the analyses of large evolving graphs. We are aiming
also to research how these representations may be used within distributed
graph processing systems in order to reduce the memory pressure
observed often in these systems.

\Section{Acknowledgments}
This research has received funding from the European Union's Horizon 2020
research and innovation programme under the Marie Sk{\l}odowska-Curie [grant
agreement No 690941], namely while the first three authors were visiting either
the University of Chile or Enxenio SL.  This work was partially funded by
Funda\c{c}\~{a}o para a Ci\^{e}ncia e a Tecnologia (FCT) [grants FCT
TUBITAK/0004/2014, CMUP-ERI/TIC/0046/2014, PTDC/CCI-BIO/29676/2017,
UID/CEC/50021/2019], by MICINN-AEI (PGE and ERDF) [grants TIN2016-77158-C4-3-R,RTC-2017-5908-7],
by Xunta de Galicia (co-founded by ERDF) [grants ED431G/01, ED431C 2017/58],
%by Xunta de Galicia/FEDER-UE [grants ED431G/01, ED431C
%2017/58], by MINECO-AEI/FEDER-UE [grants TIN2016-77158-C4-3-R,
%TIN2016-78011-C4-1-R], by MICINN [grant RTC-2017-5908-7], 
and by the Millennium Institute for Foundational Research on Data (IMFD).

\Section{References}
\bibliographystyle{IEEEbib}
\bibliography{refs}

\begin{thebibliography}{10}

\bibitem{DBLP:conf/www/BoldiV04}
Paolo Boldi and Sebastiano Vigna,
\newblock ``The webgraph framework {I:} compression techniques,''
\newblock in {\em World Wide Web Conference ({WWW})}, 2004, pp. 595--602.

\bibitem{Brisaboa2014}
Nieves~R. Brisaboa, Susana Ladra, and Gonzalo Navarro,
\newblock ``Compact representation of web graphs with extended functionality,''
\newblock {\em Information Systems}, vol. 39, pp. 152--174, 2014.

\bibitem{DBLP:journals/is/BrisaboaCBN17}
Nieves~R. Brisaboa, Ana Cerdeira{-}Pena, Guillermo de~Bernardo, and Gonzalo
  Navarro,
\newblock ``Compressed representation of dynamic binary relations with
  applications,''
\newblock {\em Information Systems}, vol. 69, pp. 106--123, 2017.

\bibitem{DBLP:conf/pods/MunroNV15}
J.~Ian Munro, Yakov Nekrich, and Jeffrey~Scott Vitter,
\newblock ``Dynamic data structures for document collections and graphs,''
\newblock in {\em {ACM} Symposium on Principles of Database Systems ({PODS})},
  2015, pp. 277--289.

\bibitem{navarro2016compact}
Gonzalo Navarro,
\newblock {\em Compact data structures: A practical approach},
\newblock Cambridge University Press, 2016.

\bibitem{quijada2019set}
Carlos Quijada-Fuentes, Miguel~R. Penabad, Susana Ladra, and Gilberto
  Guti{\'e}rrez,
\newblock ``Set operations over compressed binary relations,''
\newblock {\em Information Systems}, vol. 80, pp. 76--90, 2019.

\bibitem{DBLP:conf/spire/ArroyueloBGN19}
Diego Arroyuelo, Guillermo de~Bernardo, Travis Gagie, and Gonzalo Navarro,
\newblock ``Faster dynamic compressed d-ary relations,''
\newblock in {\em String Processing and Information Retrieval ({SPIRE})}, 2019,
  pp. 419--433.

\bibitem{BoVWFI}
Paolo Boldi and Sebastiano Vigna,
\newblock ``The {W}eb{G}raph framework {I}: {C}ompression techniques,''
\newblock in {\em World Wide Web Conference (WWW)}, 2004, pp. 595--601.

\bibitem{BRSLLP}
Paolo Boldi, Marco Rosa, Massimo Santini, and Sebastiano Vigna,
\newblock ``Layered label propagation: A multiresolution coordinate-free
  ordering for compressing social networks,''
\newblock in {\em World Wide Web Conference (WWW)}, 2011, pp. 587--596.

\bibitem{chung2003duplication}
Fan Chung, Linyuan Lu, T~Gregory Dewey, and David~J Galas,
\newblock ``Duplication models for biological networks,''
\newblock {\em {Journal of Computational Biology}}, vol. 10, no. 5, pp.
  677--687, 2003.

\bibitem{bhan2002duplication}
Ashish Bhan, David~J Galas, and T~Gregory Dewey,
\newblock ``A duplication growth model of gene expression networks,''
\newblock {\em Bioinformatics}, vol. 18, no. 11, pp. 1486--1493, 2002.

\end{thebibliography}

\end{document}